\documentclass{article}
\usepackage[preprint]{neurips_2026}

\usepackage[utf8]{inputenc}
\usepackage[T1]{fontenc}
\usepackage[hyphens]{url}
\usepackage{hyperref}
\usepackage{booktabs}
\usepackage{amsmath,amssymb}
\usepackage{microtype}
\usepackage{xcolor}
\usepackage{xspace}

\newcommand{\HUMAN}{\textsc{HUMAN}}
\newcommand{\WPR}{\textsf{W-PR}}
\newcommand{\WTR}{\textsf{W-Tran}}
\newcommand{\WSG}{\textsf{W-Sugg}}

\usepackage{enumitem}
\usepackage{array}
\usepackage{tabularx}
\usepackage{tikz}
\usetikzlibrary{positioning}
\usepackage{graphicx}
\usepackage{pifont}
\usepackage{caption}
\usepackage[framemethod=default]{mdframed}
\newcommand{\cmark}{\ding{51}}
\mdfdefinestyle{contractbox}{
  linewidth=0.6pt,linecolor=black!55,
  backgroundcolor=gray!5,
  innertopmargin=3pt,innerbottommargin=3pt,
  innerleftmargin=6pt,innerrightmargin=6pt,
  skipabove=2pt,skipbelow=2pt,
  roundcorner=2pt,
  nobreak=true,
}
\mdfdefinestyle{calloutbox}{
  linewidth=0pt,
  leftline=true,linecolor=blue!60!black,linewidth=2.5pt,
  backgroundcolor=blue!4,
  innertopmargin=2pt,innerbottommargin=2pt,
  innerleftmargin=8pt,innerrightmargin=6pt,
  skipabove=2pt,skipbelow=2pt,
}
\newcommand{\Description}[1]{}

\begin{document}
\raggedbottom
\title{Guarded Commits: Transactional Human Approvals for LLM Workflows}

\author{%
  Laurent Bindschaedler \\
  MPI-SWS \\
  \And
  Ferdi Kossmann \\
  MIT CSAIL \\
  \And
  Chunwei Liu \\
  Purdue University \\
  \And
  Jason Mohoney \\
  MIT CSAIL \\
}

\maketitle

\begin{abstract}
LLM workflows often require human approval before an irreversible external action. Most systems keep that approval outside the workflow, as an interface click or an audit entry. The workflow therefore lacks a commit-time check that every risky path reached an approval gate. Its logs may not preserve the reviewed evidence or the conditions for reusing an earlier decision. We present a guarded-commit design that makes human approval part of workflow state. Before the external action runs, a decision source appends a resolution record and its referenced evidence to a ledger. A credential-confined commit adapter then checks that record against the artifact, policy version, and executed path. Our evidence is limited to trace reconstruction. A validator test on synthetic acyclic workflow plans accepts unfaulted plans and rejects plans with each injected fault: a missing gate, the wrong gate type, or incomplete path coverage. Across three public workloads totaling 271{,}035 traces, replay reproduces recorded artifact hashes when present. Avoided reviews and disagreement with recorded decisions vary by workload. The resulting record supports audit and replay under the policy in force when the resolution was recorded.
\end{abstract}

\section{Introduction}
\label{sec:intro}

Consider an agent-authored pull request. A reviewer inspects the diff, checks logs, and approves the merge. The same pattern appears when a translator publishes a machine-translated article or a writing assistant submits a final draft. Each approval sits directly before an irreversible action. If the approval is bypassed, stale, or impossible to reconstruct, the system cannot explain why the action committed.

When approval is stored outside workflow state, the evidence a reviewer sees is not bound to the process that later commits the action. Orchestrators dispatch work while external UIs collect approvals, and logs retain the outcomes. The reviewed evidence rarely becomes a fixed workflow object, so later steps can combine previously retrieved material with new human input. Command approvals, allowlists, and audit logs remain useful operational controls~\cite{Betts2026AgentSandbox,AnthropicNDClaudeCodeSecurity}, but do not necessarily leave a record that the final commit check can validate.

Agent runtimes also separate approval from the authority to act. In pre-agent workflows, the person who approved a GitHub merge typically held the session token that authorized it, so the credential bound approval and execution. In an agent runtime, the agent process may hold the merge credential while the human reviews the diff elsewhere. The merge can then proceed without the human approval.

\textbf{Logs are not commit predicates.} Logs describe past events; they do not authorize the next action. Before executing an action, the adapter must check an approval predicate. A log saying ``Alice approved'' does not identify the artifact, execution path, or policy version covered by her approval. We therefore store each approval as a typed record in an append-only ledger. At commit, the adapter checks the ledger record against the pending action. Audit logs still support historical review, but the commit path relies on the typed record.

\textbf{Guarded-commit design.} We make each approval a workflow transaction over tracked state. The design combines an append-only approval ledger, executed-path validation, and a credential-confined commit adapter. Its contract binds each human, reused, or automated decision to the relevant evidence, policy, and execution path. Here, ``transaction'' means a commit discipline for irreversible side effects, not classical ACID execution of the agent loop. Under \emph{credential confinement} (\S\ref{sec:background}), only the adapter holds the credential for the external action. It refuses the action unless the record contains a valid resolution for the executed path under the governing policy.

\paragraph{Contributions.}
\begin{itemize}[leftmargin=*]
\item \textbf{A guarded-commit design and contract.} Current agent stacks separate command-time controls (permission prompts, allowlists, and CI checks) from target-state controls (branch protection and workflow approvals). Neither produces a workflow-state record for the commit adapter to validate. Our design records the human's decision and enforces it at the commit boundary. The contract binds that decision to the reviewed evidence, policy version, and execution path. Reusing a past decision appends a fresh record for the current context. We formalize the contract as a four-condition predicate.
\item \textbf{Trace-backed evidence.} We reconstruct the contract's records from public code-review, translation, and writing-assistance workloads. In the translation workload, the ledger identifies one bypass rule as the source of added disagreement. Because we do not compare against a deployed approval pipeline, these experiments establish feasibility rather than deployment superiority.
\end{itemize}

\section{Problem Setting and Running Example}
\label{sec:background}

\paragraph{Running example: agent-authored merge.}
A coding agent proposes a pull request that touches source files, tests, and deployment configuration. Before merge, a reviewer examines the diff, CI output, static-analysis results, and repository policies. Merging into the protected branch is the irreversible action. Publication and submission workflows follow the same pattern: a person reviews assembled evidence before the workflow performs an action that local rollback cannot cleanly undo. The merge workflow must confine its credential to a commit-validating adapter; otherwise the agent can call the merge API and bypass the approval gate. Appendix~\ref{app:credential-confinement} sketches deployment patterns. We use \texttt{ACT} for the workflow step that invokes the irreversible API. We call the blocking approval step a \emph{barrier} and write it as \HUMAN{}. The adapter checks a recorded execution path, called an \emph{executed-path witness}, to confirm that the barrier covered the pending \texttt{ACT}.

\begin{figure}[t]
\centering
\resizebox{\columnwidth}{!}{%
\begin{tikzpicture}[
  node distance=0.85cm and 1.1cm,
  box/.style={draw, rounded corners, align=center, minimum width=2.2cm, minimum height=0.95cm, font=\small, inner sep=4pt},
  humanbox/.style={box, fill=blue!15, draw=blue!60, thick},
  actbox/.style={box, fill=red!10, draw=red!60},
  ledgerbox/.style={box, fill=green!10, draw=green!60},
  >={latex},
  label/.style={font=\small\bfseries, anchor=south},
]
  \begin{scope}
    \node[box, fill=gray!15, draw=black!50] (legacy_orch) at (0,0) {Orchestrator};
    \node[box, fill=orange!20, draw=orange!70, thick, below=of legacy_orch] (legacy_ui) {External\\approval UI};
    \node[actbox, below=of legacy_ui] (legacy_act) {\texttt{ACT}};
    \draw[black!60, rounded corners, thick]
      ([xshift=-0.7cm,yshift=0.45cm]legacy_orch.north west)
      rectangle ([xshift=0.7cm,yshift=-0.45cm]legacy_act.south east);
    \node[label, anchor=south] at ([yshift=0.55cm]legacy_orch.north) {\color{black}Separate approval};
    \draw[->, thick] (legacy_orch.south) -- (legacy_ui.north);
    \draw[->, dashed, red!80, very thick] (legacy_ui.south) --
      node[right, font=\footnotesize, text=red!80] {no guarantee}
      (legacy_act.north);
  \end{scope}

  \begin{scope}[xshift=6.5cm]
    \node[box, fill=gray!15, draw=black!50] (new_retrieve) at (0,0) {\texttt{RETRIEVE}/\\\texttt{VERIFY}};
    \node[humanbox, below=of new_retrieve] (new_human) {\HUMAN{}\\barrier};
    \node[actbox, below=of new_human] (new_act) {\texttt{ACT}};
    \node[ledgerbox, right=of new_human] (ledger) {Append-only\\ledger};
    \draw[blue!70!black, rounded corners, thick]
      ([xshift=-0.7cm,yshift=0.45cm]new_retrieve.north west)
      rectangle ([xshift=0.5cm,yshift=-0.45cm]ledger.south east |- new_act.south);
    \node[label, anchor=south] at ([yshift=0.55cm]new_retrieve.north) {\color{blue!70!black}Guarded commit};
    \draw[->, thick] (new_retrieve.south) -- (new_human.north);
    \draw[->, blue!70!black, very thick] (new_human.south) --
      node[right, font=\footnotesize, text=blue!70!black, align=left] {executed-path\\witness}
      (new_act.north);
    \draw[->, green!50!black, thick] (new_human.east) -- (ledger.west);
    \draw[->, green!50!black, thick] (new_act.east) -| (ledger.south);
  \end{scope}
\end{tikzpicture}%
}
\caption{Approval outside workflow state (left) and a guarded commit (right). The ledger records \HUMAN{} decisions, and \texttt{ACT} commits only after the adapter confirms that the executed path contains the required approval.}
\Description{Two side-by-side workflow diagrams. Left: an orchestrator routes through a separate approval UI to ACT with a dashed no-guarantee edge. Right: the workflow routes retrieval and verification through a human approval before ACT, with approval and action events written to an append-only ledger.}
\label{fig:barrier-overview}
\end{figure}
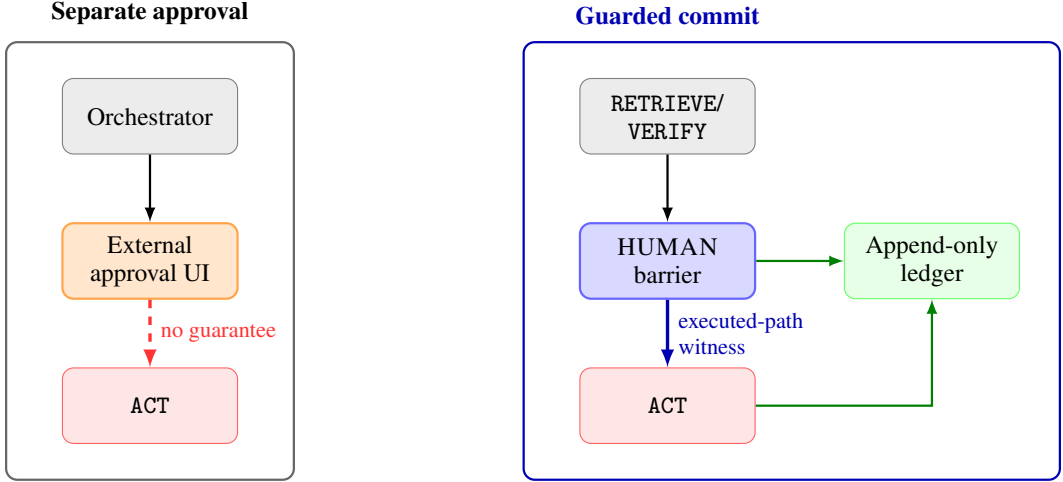

\paragraph{Approval outside workflow state.}
An LLM workflow forms a DAG connecting context retrieval and decision steps to verification and an external action. In many production deployments, an external system mediates human approval outside the orchestrator's control. The orchestrator therefore cannot require a resolved approval on every path to the action. Its log may preserve the outcome without preserving the evidence the human reviewed. Reusing a decision also requires an explicit test that the current evidence matches the evidence covered by the earlier decision. Figure~\ref{fig:barrier-overview} contrasts an external approval flow with approval recorded inside the workflow transaction.

\paragraph{Related work.}
Several research areas address parts of the approval problem. Human-in-the-loop data systems integrate human judgments into query and data processing~\cite{Franklin2011CrowdDB,Marcus2011Crowdsourced,Marcus2011Qurk,Park2012Deco,Park2013DecoOpt,Wang2012CrowdER}. Weak-supervision systems optimize human effort in model and data loops~\cite{Xin2018HelixHITL,Krishnan2016ActiveClean,Rekatsinas2017HoloClean,Bach2019SnorkelDryBell}, while selective classification and learning-to-defer address risk-controlled deferral~\cite{Geifman2017Selective,Geifman2019SelectiveNet,Madras2018Defer,Mozannar2020Defer,Lykouris2024DeferQueue}. LLM moderation research studies disagreement among human raters~\cite{Thomas2025HarmfulContentLLM,Kumar2024ContentModeration}. Software-engineering agents such as SWE-agent operate upstream of authorization: they edit repositories and run tests to solve programming tasks~\cite{Yang2024SWEAgent}. Workflow research and business-process management (BPM) have long studied approval patterns~\cite{Hollingsworth1995WorkflowReference,Aalst2003WorkflowPatterns,Aalst2013BPMSurvey,GarciaMolina1987Sagas}. Our comparison asks which commit-time clauses these mechanisms supply.

Existing agent and workflow stacks implement subsets of the contract's clauses. GitHub branch protection and Copilot cloud-agent gates bind approval to diff state, but not to a frozen evidence snapshot, policy version, or execution path. Command-time permission systems in Claude Code, Cursor, and ChatGPT Workspace gate tool calls before they fire but produce no approval record for a commit adapter to validate. CI-time policy-as-code tools such as CommitGuard and Open Policy Agent (OPA) check diffs against static rules but do not pin approval to a fixed ledger state or record a derived approval on reuse. BPM and durable-workflow engines provide process structure but not content-addressed evidence snapshots or canonical-equivalence reuse. Our contract combines an executed-path witness, a content-addressed evidence snapshot, pinning of the policy and visible ledger state, and a fresh derived record for each reuse. Appendix~\ref{app:baseline-comparison} compares these clauses directly.

\paragraph{Scope and threat model.}
The design applies to workflows with \emph{irreversible external side effects}. Other agent interactions need not undergo human review. Learned uncertainty estimation remains complementary. For an irreversible action, the system stores the approval in workflow state and checks the record against the policy under which the decision was made. We consider one decision source, or \emph{resolver}, per gate. Multi-party approvals, continuous streaming actions, and a compromised trusted base are out of scope.

The trusted base is the transaction manager, plan validator, \texttt{ACT} adapter, and ledger append service, all assumed honest. The agent process and ordinary orchestration logic are not trusted to authorize commits. The design presupposes credential confinement: the credential that authorizes the irreversible action must be held by the \texttt{ACT} adapter rather than by the agent process. Whether a given agent runtime enforces this boundary is a deployment question (Appendix~\ref{app:credential-confinement}) we do not solve. IDE-based agent stacks such as Claude Code and Cursor~\cite{AnthropicNDClaudeCodeSecurity,Betts2026AgentSandbox} sandbox filesystem and network access, but typically run the bash or terminal tool with the user's shell credentials reachable, so an agent that wants to bypass the commit adapter has the credential to do so. An attacker who controls the agent process or the orchestrator can prevent commits but cannot forge an approval record or bypass the adapter if credentials are confined. Control of any trusted-base component defeats the guarantee.

\section{Guarded-Commit Design}
\label{sec:substrate}

The guarded-commit design makes human approval enforceable at commit. An append-only ledger stores the evidence snapshot and decision, while a plan validator establishes which barriers each execution path requires. Before running the irreversible action, the \texttt{ACT} adapter checks the record against the artifact, policy, and executed path. A failed check blocks the action and routes the decision to human review.

A resolver supplies the decision as a human judgment, a validity-checked reuse of a past decision (\S\ref{subsec:knowledge-retrieval}), or a policy-authorized automation rule (\S\ref{subsec:module-optimization}). The decision source changes the frequency of human review but not the commit check. Each source must append a decision record and its referenced evidence before the adapter invokes the external API. Every decision record must match the transaction's pinned state.

\paragraph{Transaction scope.}
For one guarded action, a workflow transaction fixes its inputs at defined points. The ledger prefix is the ledger state visible when the transaction begins. The transaction adds a commit check before external \texttt{ACT}s but provides no atomic rollback, two-phase locking, or snapshot isolation for arbitrary tool state. A separate \emph{transaction manager} assigns the transaction id and pins the governance-policy version and ledger prefix at the start. It fixes the plan fragment before validation and the approved artifact when the resolution record is appended. Only the transaction manager can advance the run to a new transaction state. The orchestrator and \texttt{ACT} adapters can read the fixed values but cannot change them. The fixed plan and policy provide the reference for plan validation; the full set provides the reference for the commit check.

\paragraph{Plan and \HUMAN{} barrier.}
A plan states where approval is required. The plan is a DAG of typed steps (\texttt{RETRIEVE}, \texttt{VERIFY}, \HUMAN{}, \texttt{ACT}); an \texttt{ACT} has an external side effect that local rollback cannot undo. A \HUMAN{} barrier's \emph{type} specifies the protected action, required reviewer capability, permitted decisions (\texttt{APPROVE}, \texttt{EDIT} with delta $\Delta$, or \texttt{REJECT}), and evidence schema. The reviewer sees a \emph{context snapshot}, the deterministic evidence bundle required by that schema. The barrier records the snapshot, reviewer identity and credentials, policy version, and timestamps. The barrier type and snapshot define the approval record expected at commit.

The \texttt{ACT} adapter checks barrier placement against the executed path. A \emph{barrier-coverage certificate} names the barriers that the validator requires for the action. An \emph{executed-path witness} lists the validated plan nodes and edges traversed from the start to \texttt{ACT}. The transaction-managed runtime emits the witness; alternatively, the validator reconstructs the path from signed step events. The LLM never supplies the path. The validator rejects a witness that omits a step declared by the validated DAG. The review UI must render logged, content-addressed artifacts and record the renderer and schema versions so the context can be replayed (Appendix~\ref{app:replay-modes}). Ledger offsets increase monotonically: the manager reads at the pinned offset and appends the decision after that offset, and the adapter may consume the decision only once per transaction. The certificate and witness therefore show that the executed path crossed an eligible barrier.

\paragraph{Policy roles.}
Two policies separate what approval is required from how the workflow obtains approval. The \emph{governance policy} maps actions and cohorts to barrier types, reviewer capabilities, evidence schemas, verifier versions, credential-freshness rules, and permitted automation. The governance policy also fixes the version of the canonicalizer used to produce normalized comparison keys. The \emph{resolution policy} (\S\ref{subsec:module-optimization}) chooses, for each barrier, between human review, decision reuse, an automation rule, and escalation. Implementations may replace the resolution policy without changing the commit predicate. At transaction start, the manager pins the governance-policy version and ledger prefix. Coverage, credential, and reuse checks use those pinned values and cannot cross a policy or ledger change within the transaction.

\paragraph{Safety invariant.}
Barrier coverage is the central safety property. Coverage takes two forms: graph-theoretic dominance when one barrier suffices, and a branch-specific barrier cut when different branches need different barrier types. The commit predicate combines coverage with checks on the pinned state, artifact, and source of the decision.

\begin{mdframed}[style=contractbox]
\textbf{Guarded-commit contract.} At commit, the adapter evaluates the current transaction's barrier-resolution record under the governance-policy version and ledger prefix fixed at transaction start. It invokes \texttt{ACT} only if all four conditions hold.
\begin{enumerate}[leftmargin=18pt, itemsep=1pt, topsep=2pt]
  \item \textbf{Barrier coverage with path obligations.} Every validated path from the plan start to the \texttt{ACT} carries a \emph{path obligation} (path predicate, required barrier type, eligible barrier identities). The executed-path witness must be a path in the validated DAG, must match exactly one declared obligation, and must contain a resolved barrier whose identity and type are eligible for that obligation. A barrier resolved on a different path does not satisfy coverage for this one.
  \item \textbf{Policy and knowledge pinning.} The resolver produced the resolution under the same (governance-policy version, ledger prefix) the transaction pinned.
  \item \textbf{Artifact identity (with \texttt{EDIT} triple).} The artifact consumed by \texttt{ACT} matches the artifact hash bound into the resolution. For \texttt{EDIT}, deterministic application of $\Delta$ to $H_\mathrm{pre}$ produces $H_\mathrm{post}$ before the record is appended. The resolution binds the triple $(H_\mathrm{pre}, H_\Delta, H_\mathrm{post})$, and \texttt{ACT} consumes $H_\mathrm{post}$. Any mutation after the record is appended invalidates it.
  \item \textbf{Safe reuse / automation.} For decision-cache reuse: canonical contexts equivalent under the current canonicalizer; prior reviewer's credentials satisfy the current barrier under the policy-declared freshness test; the reuse record resolves transitively to a non-derived \HUMAN{} or policy-declared automation root in the pinned prefix; the workflow appended a fresh derived resolution record (\S\ref{subsec:knowledge-retrieval}). For an automation rule: the rule's declared verifier checks pass (\S\ref{subsec:module-optimization}).
\end{enumerate}
If any condition fails, the adapter blocks the action. A plan or path failure returns to validation; an absent or invalid resolution returns to the resolution policy, which may request fresh \HUMAN{} review.
\end{mdframed}

Condition 1 checks the validated plan and the executed path. Conditions 2--4 run at commit against the append-only ledger, pinned policy and ledger state, and the fresh record created for any reuse.

\paragraph{Failure semantics.}
The adapter blocks an \texttt{ACT} that lacks a matching record, but a valid record does not guarantee delivery to the external system. If the API fails after a ledger append, the adapter retries with an idempotency key derived from the resolution-record hash. If the API succeeds without an append, the next reader sees no commit, so the workflow blocks further \texttt{ACT}s on the same target or requests fresh review. Exactly-once external effects depend on the adapter's idempotency contract.

\paragraph{Trace replay.}
Each \HUMAN{} event stores a content-addressed snapshot and the renderer/schema version used to present the snapshot. Artifact-level replay checks the recorded blobs and presentation contract, but cannot recover evidence shown outside the snapshot. Appendix~\ref{app:replay-modes} defines four replay modes. Freezing the recorded decision context also permits local counterfactual evaluation of different resolution policies.

\subsection{Decision Reuse}
\label{subsec:knowledge-retrieval}

A decision cache is a validity-checked index over the ledger, keyed by canonical context. The cache avoids redundant human review only when the current context is equivalent to one already adjudicated in the pinned ledger prefix, under a three-part test:
\begin{enumerate}[leftmargin=*, itemsep=0pt]
  \item \textbf{Policy match:} the governance-policy version must be the same.
  \item \textbf{Context equivalence:} a workload-specific canonicalizer maps raw contexts to canonical forms, and reuse requires that the canonical forms match.
  \item \textbf{Capability sufficiency with credential freshness:} the prior reviewer's credentials must satisfy the current barrier requirement under a policy-declared freshness test. Acceptable forms are a currently valid capability proof, a time-bounded credential assertion whose validity interval covers the reuse transaction, or a revocation check performed at reuse time for the relevant credential interval. The reuse record logs the proof source, check time, TTL if applicable, and revocation result.
\end{enumerate}
If any condition fails, the system invokes \HUMAN{} again.

Each workload defines a conservative canonical key (\S\ref{sec:workloads}). \WPR{} uses repository + governance-policy version + verifier-output digest + normalized diff; \WTR{} uses source segment + language pair + policy version; \WSG{} uses prior-text window + prompt template + policy version. The resulting keys require exact matches over normalized representations rather than semantic nearest-neighbor searches. Appendix~\ref{app:canonicalizers} gives the normalization rules and verifier-output digest.

\begin{mdframed}[style=contractbox]
\textbf{Reuse writes a new record.} A cache hit creates a new authorization event rather than redirecting the action to an old event. The fresh record binds the current context-snapshot hash, artifact hash, governance-policy version, and ledger prefix. The record also contains the prior decision's ledger offset, the canonicalizer version that established equivalence, and the credential-freshness result. Its pointer must resolve, transitively, to a non-derived \HUMAN{} or policy-declared automation root within the pinned prefix; a derived record cannot source another derived record. The \texttt{ACT} adapter validates this new record as if a human had just resolved the barrier. The equivalence proof links the new record to the prior decision.
\end{mdframed}

These exact keys are purposefully conservative. A cache hit is safe only when the key contains every decision-relevant input or a verifier covers any omitted input. Within the reuse path, a cache miss returns the decision to human review; it never authorizes the action. At a 0\% exact-match rate, reuse saves no labor, while the commit check, audit record, and replay guarantees remain unchanged. 

\subsection{Optional Automation}
\label{subsec:module-optimization}

The optional automation layer records why the workflow invoked human review, including failed checks, applicable rules, and the resulting decision. A separate difference log records changes between system proposals and human approvals or edits. The resolution policy may use the log to reduce human review and measure disagreement with recorded decisions. Neither record is required by the commit predicate. Disagreement is a proxy for review outcome, not a measure of correctness.

\paragraph{Failure attribution.}
These records make three failure modes easier to trace. \emph{Over-broad canonicalization} maps different contexts to the same key and can apply an earlier decision where it does not belong. A \emph{misplaced barrier} leaves a runtime path outside the validated plan, so the supplied DAG appears covered while execution is not. \emph{Over-broad automation} applies a rule beyond the cases checked by its verifier. The ledger records the key, path, and rule used for each resolution, which helps isolate the mechanism behind a disagreement. The \WTR{} result in \S\ref{subsec:rq-optimization} illustrates the third case.

\paragraph{Automation rules.}
An avoidance rule is a versioned fast path, so the policy must still define its safe scope. The rule may suppress a future invocation only when the policy declares its cohort, verifier coverage, unchecked failure modes, and rollback or reversion procedure. A ``docs-only auto-merge'' rule, for instance, might bypass \HUMAN{} only when the modified files belong to a predefined allowlist and all secondary validation checks pass. Changes touching authentication, cryptography, dependency manifests, or privileged deployment scripts remain ineligible. A parallel rule for \WTR{} bypasses review only when the segment is shorter than a length cap, glossary and structural checks pass, and the source domain belongs to a predeclared low-risk cohort. Declared coverage, not observed agreement, makes the rule auditable.

\paragraph{Learning from reviewer corrections.}
The difference log pairs system proposals with human approvals or edits. Recurring fixes, such as terminology corrections in translation or formatting changes in pull-request review, can become automated \texttt{VERIFY} checks. A new automation rule must still declare its scope and verifier coverage; observed agreement alone does not authorize it.

\paragraph{Resolution policy.}
Given a barrier and the pinned governance policy, the resolution policy chooses human review, decision-cache reuse, a policy-authorized automation rule, or escalation. The ledger records the choice, the policy inputs, and the policy version. Our evaluation sweeps three configurations: \textbf{Fresh review}, \textbf{Cache only}, and \textbf{Cache + rules}. Learning-to-defer methods can replace the resolution policy without changing the commit predicate if they choose from the same actions.

\section{Workloads}
\label{sec:workloads}

The workloads apply the same contract to different review decisions. Table~\ref{tab:workload-data} summarizes public traces for code review, translation, and writing assistance. \WPR{} (AIDev-pop~\cite{Li2026AIDev}) captures agent-authored GitHub PRs reviewed before merge to a protected branch. Its canonical key is the tuple in \S\ref{subsec:knowledge-retrieval}, and an automation rule enables a docs-only bypass. \WTR{} (Wikimedia ContentTranslation~\cite{Laxstrom2015ContentTranslation}) records MT drafts post-edited section-by-section before publication. Semantic adequacy is broader than exact diff equivalence, which makes it the hardest workload for conservative reuse. \WSG{} (CoAuthor~\cite{Lee2022CoAuthor}) records keystroke-level accept/reject/edit on writing suggestions before submission. Its stakes are lower, but its replay signal is finer-grained. The workloads therefore span distinct review contexts while sharing one commit predicate.

\begin{table}[t]
\centering
\caption{Public trace workloads used in the evaluation. A\,=\,approve, R\,=\,reject, E\,=\,edit; \cmark\,=\,present, --\,=\,absent. None includes native credential proofs.}
\label{tab:workload-data}
\small
\begin{tabular}{@{}lrcccc@{}}
\toprule
 & Traces & Decisions & Time & $\Delta$ & Credential proofs \\
\midrule
\WPR{}  & 33{,}056  & A/R   & \cmark & --         & -- \\
\WTR{}  & 236{,}532 & A/E   & --         & \cmark & -- \\
\WSG{}  & 1{,}447   & A/E/R & \cmark & \cmark & -- \\
\bottomrule
\end{tabular}
\end{table}

\section{Evaluation on Trace-Backed Workloads}
\label{sec:evaluation}

We measure avoidance and disagreement under conservative reuse and rule-based bypass, with separate validator and replay-hash checks. Because the public traces lack native reviewer credential proofs, cache-only avoidance measures \emph{context-recurrence reuse potential} under conservative canonicalization, not validated reuse under the full credential-checked contract. Each counterfactual uses a frozen decision context so that only the resolution-policy configuration changes.

\subsection{Methodology}
\label{subsec:methodology}

We evaluate the three workloads from \S\ref{sec:workloads}: \WPR{} (33{,}056 traces), \WTR{} (236{,}532), and \WSG{} (1{,}447). The conversion pipeline assigns an explicit governance-policy version to every converted trace. We restrict counterfactual conclusions to fixed recorded contexts. The conversion is mechanical: \WPR{} uses normalized PR diffs and verifier-output digests, \WTR{} uses the source segment and language pair, and \WSG{} uses the prior-text window and prompt template. The active automation rules are the docs-only PR rule and low-risk translation rule sketched in \S\ref{subsec:module-optimization}; \WSG{} has no active rule. The workload mappings supply the fixed inputs for the policy sweep.

\paragraph{Baselines and metrics.}
The policy sweep compares Fresh review, Cache only, and Cache + rules (\S\ref{subsec:module-optimization}). We process records chronologically for \WPR{} and \WSG{}, where timestamps exist, and use dump order as a chronology proxy for \WTR{}. Only past decisions populate the cache; we permit no look-ahead reuse. This ordering keeps each reuse decision within the evidence available at that point in the trace.

\textbf{Fresh review} is the operational baseline for the sweep. We do not compare against a live deployed pipeline, such as GitHub branch protection plus agent gates, or against a learned deferral baseline~\cite{Madras2018Defer,Mozannar2020Defer,Lykouris2024DeferQueue}. The experiments therefore test internal failure-mode attribution and auditability, not performance against deployed pipelines. Headline metrics are validator accuracy (correct acceptance of unfaulted synthetic plans and rejection of faulted ones), artifact coverage (the fraction of traces with recorded artifacts), replay-hash equality among traces with artifacts (\S\ref{subsec:rq-safety}), avoidance rate (the fraction of reviews skipped), and disagreement rate (the fraction of resolution-policy decisions that differ from recorded human decisions). We report disagreement over all traces and among avoided cases. Capability-check numbers remain synthetic predicate sanity checks rather than measurements of real authorization failures.

\begin{figure}[t]
\centering
\begin{minipage}[t]{0.6\columnwidth}
\vspace{0pt}%
\centering
\scriptsize
\setlength{\tabcolsep}{3pt}
\begin{tabular}{llrrr}
\toprule
Group & Metric & \WPR{} & \WTR{} & \WSG{} \\
\midrule
\multicolumn{5}{l}{\textit{Plan validator (synthetic DAGs)}} \\
& Validator accuracy$^\dagger$ & \multicolumn{3}{c}{1.000 (999/999 synthetic DAGs)} \\
\multicolumn{5}{l}{\textit{Artifact-hash round-trip}} \\
& Artifact coverage          & 100\% & 92.1\% & 100\% \\
& Replay match              & 1.000 & 1.000 & 1.000 \\
\multicolumn{5}{l}{\textit{Resolution-policy sweep}} \\
Cache only & Avoidance         & 0.9\% & 16.8\% & 72.0\% \\
           & Disagreement      & 0.04\% & 5.3\% & 27.6\% \\
Cache + rules & Avoidance      & 8.5\% & 25.0\% & 72.0\% \\
              & Disagreement   & 0.1\% & 10.6\% & 27.6\% \\
\multicolumn{5}{l}{\textit{Dataset audit}} \\
& Gov.-policy version assigned & 100\% & 100\% & 100\% \\
& Native credential proof present & 0\% & 0\% & 0\% \\
\bottomrule
\end{tabular}
\captionof{table}{Main trace-backed results. $^\dagger$Validator accuracy is the accept/reject rate on synthetic faulted and unfaulted DAGs, not a per-workload bypass rate. Replay match is hash equality among traces with recorded artifacts; public traces contain no native credential proofs.}
\label{tab:main-results}
\end{minipage}\hfill
\begin{minipage}[t]{0.38\columnwidth}
\vspace{0pt}%
\centering
\includegraphics[width=\linewidth,trim=0 0 0 8pt,clip]{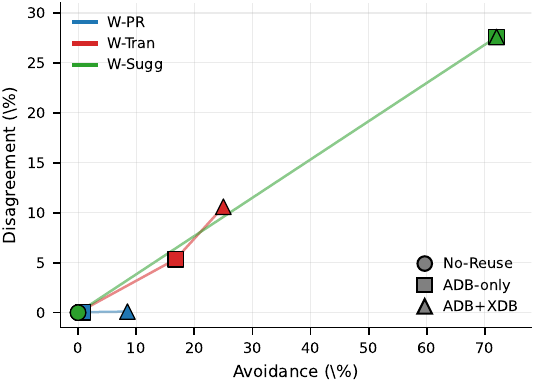}
\captionof{figure}{Avoidance--disagreement curve for Fresh review, Cache only, and Cache + rules. Connected points are sweep configurations, not optimized Pareto points.}
\label{fig:frontier}
\end{minipage}
\end{figure}

\subsection{Avoidance--Disagreement Curve}
\label{subsec:rq-retrieval}

Table~\ref{tab:main-results} reports the validator, artifact-hash round-trip, sweep, and dataset-audit results. Figure~\ref{fig:frontier} plots the avoidance--disagreement curve. Cache-only avoidance ranges from 0.9\% (\WPR{}) to 72.0\% (\WSG{}), with overall disagreement ranging from 0.04\% to 27.6\%. Recurrence and avoidance match to within rounding, which separates canonicalizer limits from workload repetition.

Adding rules raises avoidance to 8.5\% / 25.0\% / 72.0\%. \WPR{} gains at $\sim$0.1\% overall disagreement, \WSG{} gains nothing, and \WTR{} doubles overall disagreement to 10.6\%. Conditional disagreement among avoided reviews is $\sim$1\% for \WPR{} with cache and rules, $\sim$42\% for \WTR{} with cache and rules, and $\sim$38\% for \WSG{} with cache alone. The \WTR{} and \WSG{} configurations are diagnostic stress points rather than deployable bypass policies. Here, disagreement means divergence from the recorded review outcome. Severity differs by workload: \WPR{} and \WSG{} decisions are largely reversible, while \WTR{} publishes externally visible content. Disagreement does not bound severity-weighted loss in either direction; severity-weighted analysis belongs at deployment.

\subsection{Attributing the \WTR{} Disagreement Spike}
\label{subsec:rq-optimization}

We use the \WTR{} Cache + rules configuration to attribute disagreement to one bypass rule. Cache only contributes 16.79\% avoidance at 5.35\% overall disagreement. The rule adds 8.21 percentage points of avoidance and 5.20 percentage points of overall disagreement. The bypassed cohort therefore has roughly 63\% conditional disagreement (5.20 / 8.21). The combined configuration has roughly 42\% conditional disagreement among all avoided reviews. In isolation, the rule is an unsafe class-level bypass. Ledger entries identify the rule as the source of the added disagreements.

A keyword-pattern audit examines the 25{,}004 disagreements among cases avoided by the cache-and-rules configuration. The set includes all avoided-case disagreements, not only the $\sim$12{,}300 attributable to the rule. The audit labels 3{,}444 cases (13.8\%) with three failure modes that the rule's declared verifier coverage (glossary adherence, segment length, and structural consistency) does not detect: \emph{disfluency} (1{,}768 cases where humans smoothed MT output that passed the verifier), \emph{reordering} (1{,}531 cases where humans moved phrases), and \emph{terminology} (145 cases of domain-specific word choice). The remaining 86.2\% are unclassified. Because we did not run a manual precision/recall audit, these labels are illustrative examples rather than a taxonomy.

\paragraph{Interpretation.}
The \WTR{} rule applies beyond the safe scope declared in \S\ref{subsec:module-optimization}. The ledger supports internal attribution, but the experiment does not show that a deployed pipeline would miss the same failures.

\subsection{Contract Checks}
\label{subsec:rq-safety}

\paragraph{Coverage-validator unit test.}
We ran the coverage validator on 999 author-generated synthetic acyclic plan DAGs across 333 branching topologies, with three injected fault classes: missing barrier, wrong-kind barrier, and partial-coverage barrier. The validator produced 0 false accepts and 0 false rejects. The measurement is a coverage-algorithm unit test under the injected fault model. Cyclic plans (retry loops and alternate-reviewer loops) must be unrolled into finite DAGs for each transaction. The test does not measure real-world bypass resistance, credential confinement, or \texttt{ACT}-adapter security.

\paragraph{Replay-hash fidelity.}
For traces with recorded artifacts, Playback replay reproduced the recorded content hash in 252{,}386/252{,}386 cases (artifact coverage 100\% / 92.1\% / 100\%, where the 7.9\% \WTR{} gap is null source-dump targets). The replay-hash result verifies content-addressed artifact reconstruction only; it does not test live-endpoint replay or reviewer-UI fidelity.

\paragraph{Synthetic capability-predicate sanity check.}
We hash trace ids into a 75/10/10/5 valid/missing/wrong-role/expired distribution and verify that the capability-sufficiency predicate blocks every state except \emph{valid}. Block rates (30.0\% / 28.9\% / 16.7\%) are artifacts of the synthetic assignment. The predicate blocked every synthesized invalid state; these rates do not represent real-world authorization failures (Appendix~\ref{app:credential-stress}).

\section{Discussion}
\label{sec:alternatives}

\paragraph{Command-time controls.}
Command-time controls do not replace commit-time approval. Cursor and Claude Code already use permission prompts, allowlists, and audit logs to gate side effects when commands run~\cite{Betts2026AgentSandbox,AnthropicNDClaudeCodeSecurity}, and the guarded-commit design retains those controls. They do not, however, store approval as workflow state, so their logs alone cannot reconstruct what the approver saw or which policy version was in force. A barrier also becomes unenforceable when the agent process holds a credential broad enough to act directly. Commit-time coverage therefore depends on routing the action through a credential-confined \texttt{ACT} adapter (\S\ref{sec:background}).

\paragraph{Relation to BPM and durable workflows.}
BPM research has long studied approvals, routing, and process structure~\cite{Hollingsworth1995WorkflowReference,Aalst2003WorkflowPatterns,Aalst2013BPMSurvey}. The guarded-commit contract covers human decisions over dynamically assembled LLM context, including retrieval results, model outputs, tool traces, and intermediate artifacts. The ledger preserves those decisions for replay, and the adapter guards only irreversible actions. An ordinary workflow engine can host these checks without providing ACID execution for the agent loop (\S\ref{sec:substrate}).

\paragraph{Relation to uncertainty gating and learning-to-defer.}
Deferral methods~\cite{Geifman2017Selective,Geifman2019SelectiveNet,Madras2018Defer,Mozannar2020Defer,Lykouris2024DeferQueue} may match or improve the avoidance--disagreement curves reported here. They choose when to defer; they do not by themselves record what the reviewer saw or the policy version used at commit. A deferral method can therefore serve as the resolution policy while the commit check retains barrier coverage, ledger-prefix pinning, and capability-checked reuse (\S\ref{subsec:module-optimization}).

\paragraph{Limitations and open empirical questions.}
The contract establishes that an action is covered by a policy-valid resolution over recorded evidence; it does not establish that the reviewer, policy, or automation rule made the correct decision. The guarantee also depends on complete evidence schemas and an adapter that cannot be bypassed. The contract adds little when an existing pipeline already provides equivalent evidence snapshots, policy and ledger pinning, derived-approval reuse, and attribution for bypass-rule disagreements. Coverage rules that exclude safe automation need revision. A deferral method that supplies replayable approval state and commit validation without explicit coverage semantics would weaken the case for this design.

\section{Conclusion}
\label{sec:conclusion}

Guarded commit makes human approval enforceable workflow state. Before an
irreversible action, the adapter checks path coverage, policy and ledger-prefix
pinning, artifact identity, and the validity of any reused or automated
decision (\S\ref{sec:substrate}). Public traces contain enough recorded
contexts and artifacts to reconstruct the workload-side records when those
artifacts are present; plan validation and capability checks remain synthetic.
We have not yet compared deployment performance or severity-weighted outcomes
(Appendix~\ref{app:research-directions}). Within that scope, the design
specifies the state and checks required before an irreversible action commits.

\bibliographystyle{plainnat}
\bibliography{main}

\clearpage
\appendix
\section*{Appendix}
\addcontentsline{toc}{section}{Appendix}

\section{Baseline Coverage of the Four Commit-Time Clauses}
\label{app:baseline-comparison}

Table~\ref{tab:baseline-comparison} compares existing systems against four commit-time clauses: executed-path coverage, policy and ledger pinning, artifact identity, and a derived reuse record. None of the agent stacks, policy-as-code tools, or workflow engines we surveyed combines all four clauses out of the box. Configuring BPM or Temporal engines may close individual gaps. The guarded-commit layer enforces all four clauses over one ledger.

\begin{table}[h]
\centering
\caption{Representative baseline support for the four commit-time clauses. C\textsubscript{1}: executed-path witness; C\textsubscript{2}: policy/ledger pinning; C\textsubscript{3}: artifact binding; C\textsubscript{4}: derived reuse record. \cmark\ means supported by design, $\circ$ means partial or configurable, and -- means not provided.}
\label{tab:baseline-comparison}
\small
\begin{tabular}{@{}p{0.62\linewidth}cccc@{}}
\toprule
System & C\textsubscript{1} & C\textsubscript{2} & C\textsubscript{3} & C\textsubscript{4} \\
\midrule
GitHub branch-protection + Copilot agent & $\circ$ & -- & $\circ$ & -- \\
Claude Code / Cursor permission systems~\cite{Betts2026AgentSandbox,AnthropicNDClaudeCodeSecurity} & -- & -- & -- & -- \\
CommitGuard / OPA-style policy-as-code & $\circ$ & -- & $\circ$ & -- \\
BPM / Temporal-style durable workflow~\cite{Hollingsworth1995WorkflowReference,Aalst2003WorkflowPatterns,Aalst2013BPMSurvey} & $\circ$ & $\circ$ & -- & -- \\
SWE-agent software-engineering agent~\cite{Yang2024SWEAgent} & -- & -- & -- & -- \\
ChatGPT Workspace Agent action controls & -- & -- & -- & -- \\
\textbf{Guarded-commit layer} & \textbf{\cmark} & \textbf{\cmark} & \textbf{\cmark} & \textbf{\cmark} \\
\bottomrule
\end{tabular}
\end{table}

\section{Guarded-Commit Implementation Details}
\label{app:substrate-details}

\paragraph{Replay modes.}
\label{app:replay-modes}
We define four replay modes that progressively relax determinism: \emph{Playback} renders the stored trace as-is with no re-execution; \emph{Tool-Stub} re-runs workflow steps but uses recorded outputs for external calls; \emph{Retrieval-Stub} re-executes retrieval live and stubs remaining external calls; \emph{Live Re-run} re-executes everything against live endpoints. We evaluate Playback. Tool-Stub and Retrieval-Stub match Playback in trace-backed mode because they consume immutable content-addressed blobs. Live Re-run requires live endpoints and is outside this paper's evaluated scope.

\paragraph{Canonicalizer construction.}
\label{app:canonicalizers}
For \WPR{}, the canonical key is (repository, governance-policy version, verifier-output digest, normalized diff). The normalized diff is a canonical patch representation: stable file ordering, normalized line endings and metadata, deterministic hunk identifiers, and policy-declared treatment of whitespace (preserved unless the workload's governance policy declares it decision-irrelevant for the file type). The verifier-output digest hashes the CI summary and static-analysis results that the reviewer is expected to inspect, so reuse fires only when verifier evidence is also bytewise equivalent. For \WTR{}, the key is (source segment, language pair, governance-policy version). For \WSG{}, the key is (prior-text window, prompt template, governance-policy version). The keys are sound only when the workload's governance policy declares all decision-relevant evidence either included in the canonical key or covered as a verifier precondition. The decision cache also stores the full prior record so the barrier UI can display the original review context when a match is found.

\paragraph{Capability-gated reuse synthesis.}
\label{app:credential-stress}
Public traces lack native credential proofs, so we synthesize a per-trace credential state by hashing the trace identifier into one of four buckets: \emph{valid} (75\%), \emph{missing} (10\%), \emph{wrong-role} (10\%), \emph{expired} (5\%). The synthesis is uniform across workloads. The capability-sufficiency predicate (\S\ref{subsec:knowledge-retrieval}) treats anything other than \emph{valid} as insufficient and blocks reuse. Block rates (30.0\% / 28.9\% / 16.7\% for \WPR{} / \WTR{} / \WSG{}) differ from the underlying 25\% miscredential rate because reuse candidates are not uniformly distributed across credential states once the cache fills with the first-seen credential per canonical key. The synthetic rates test whether the predicate rejects invalid credentials. They do not estimate authorization-failure rates in the underlying datasets. A workload-aware credential schema (CODEOWNERS-style roles for \WPR{}, language-pair authorization for \WTR{}) is left to deployment.

\section{Deployment Patterns and Open Problems}
\label{app:deployment-and-open-problems}

\paragraph{Credential confinement deployment patterns.}
\label{app:credential-confinement}
Credential confinement makes the commit predicate enforceable by keeping the irreversible-action credential in the \texttt{ACT} adapter rather than the agent process. The credential may be a GitHub merge token, a CMS publish key, or a submission API secret. Deployment can enforce this boundary with capability-scoped tokens, network-layer egress restrictions, or OS sandboxing. With capability-scoped tokens, the transaction manager mints a short-lived token bound to the (workflow-transaction id, protected \texttt{ACT} node, artifact pre-hash) tuple, and only the adapter holds it. With network-layer restriction, agent processes cannot reach merge, publish, or submit endpoints directly; a commit-validating proxy forwards calls only after validating a barrier-resolution record. With OS sandboxing, agent tools have no filesystem or network access to the credential and expose a single RPC for requesting \texttt{ACT}. A compliant deployment leaves no path to the irreversible API outside the adapter, and a production system may combine these mechanisms.

\paragraph{Open problems.}
\label{app:research-directions}
Evidence selection for human reviewers remains an open causal problem. Context selection, ordering, and density could be learned from the ledger, but the ledger records only what reviewers saw and decided, not what they would have decided under a different presentation.

Cost-aware deferral requires a richer action space than the present rule sweep. A policy could choose among automate, reuse, route, and escalate while accounting for failure costs, queueing dynamics, and SLA bounds~\cite{Madras2018Defer,Mozannar2020Defer,Lykouris2024DeferQueue}. A direct comparison would test the rule-based selector against learned deferral under the same commit predicate.

Safe non-exact matching could increase reuse beyond exact hashing. Candidate methods include AST normalization, bounded delta allowlists, and compound model-based verifiers. Any such method must preserve the artifact and context guarantees in the commit predicate.

Sound canonicalization and replay remain separate technical problems. For canonicalization, the target is the smallest semantic equivalence that improves \WTR{} or \WSG{} reuse without raising disagreement. For replay, the corresponding problem is live-endpoint execution under nondeterministic tools.

Quorum approvals require a broader barrier model. A single-resolver barrier can extend to $k$-of-$n$ approvals or role-disjoint cuts (code-owner $\land$ security-reviewer $\land$ legal) over multiple typed barrier nodes. The four-condition predicate generalizes, but the resolution policy must address reviewer independence, conflicts of interest, cascading escalation, and multi-resolver canonicalization.

Comparative measurement requires a deployed baseline. A direct test would compare a guarded-commit deployment with a separate approval pipeline, such as GitHub branch protection plus Copilot cloud-agent gates, on the same workload. The comparison should measure attribution recall on rule-bypass disagreements, time to attribute post-publish reverts, and credential-validated reuse rate.

\end{document}